\documentclass[a4paper,fleqn]{cas-dc}
\usepackage[authoryear,longnamesfirst]{natbib}
\pdfoutput=1

\usepackage{fp} 
\usepackage{bm} 
\usepackage{amssymb} 

\renewcommand*{\Pi}{{\varPi}}

\newcommand{\N}{\mathbb{N}}
\renewcommand{\P}{\mathbb{P}}
\newcommand{\X}{\mathbf{X}} 
\newcommand{\cs}{\textbf{S}} 
\newcommand{\bfP}{\textbf{P}}  

\newcommand{\D}{DT(\textbf{sh}(t))} 

\usepackage{amsbsy} 
\usepackage{graphicx} 
\usepackage{algorithm,algorithmic}
\usepackage{bm}
\usepackage{array}
\usepackage{mathtools}
\usepackage[mathscr]{eucal}
\usepackage{listings}
\usepackage{nopageno}

\usepackage{geometry}
\begin{document}
\pagestyle{plain}

\title{Randentropy: a software to measure inequality in random systems}

\author[1]{Guglielmo D'Amico}
\cormark[1]
\ead{g.damico@unich.it}

\author[2]{Stefania Scocchera}

\author[1]{Loriano Storchi}
\cormark[2]
\ead{loriano@storchi.org}
\ead[url]{https://www.storchi.org/}

\address[1]{Department of Economics, University G. D'Annunzio Chieti - Pescara, viale Pindaro, 42 - 65127 Pescara, Italy}
\address[2]{Dipartimento di Farmacia, Universit\'a degli Studi G. D’Annunzio, 66100 Chieti, Italy}

\maketitle

\begin{abstract}
The software Randentropy is designed to estimate inequality in a random system where several individuals interact moving among many communities and producing dependent random quantities of an attribute. The overall inequality is assessed by computing the Random Theil's Entropy. Firstly, the software estimates a piecewise homogeneous Markov chain by identifying the changing-points and the relative transition probability matrices. Secondly, it estimates the multivariate distribution function of the attribute using a copula function approach and finally, through a Monte Carlo algorithm, evaluates the expected value of the Random Theil's Entropy. Possible applications are discussed as related to the fields of finance and human mobility.\end{abstract}


\section{Introduction}

 The issue of measuring inequality in a system found extensive treatment in the literature. One interesting approach is based on entropic measures. Starting from the pioneering work by \cite{shannon} on the mathematical theory of communication, the concept of entropy has found a rapid development and diffusion in many scientific communities. Notable examples are statistics (see, e.g. \cite{kullback1951information}), statistical mechanics (see, e.g. \cite{jaynes1957information}), economy (see, e.g. \cite{theil}) and ecology (see, e.g. \cite{phillips2006maximum}) just to name a few.\\   
Recent efforts have been dedicated mainly to introduce new entropies as the cumulative residual entropy (see, \cite{rao2004cumulative}) or the cumulative past entropy (see, \cite{di2009cumulative}). In the meantime, and mainly motivated by economic problems, the notion of random entropy has emerged in terms of a normalization of a random process. The random entropy shares the same functional form as the classical entropy but is related to a random process (\cite{d2010generalized}). This more general entropy was called by the author Dynamic Theil Entropy. Nevertheless we refer to it as Random Entropy, to avoid any possible misunderstanding with other dynamic entropies which are expressed as deterministic functions as in \cite{di2002entropy}, \cite{asadi2007dynamic} and \cite{cali2020properties}.

The Random Entropy allows to quantify uncertainty in a random system evolving in time and encompasses recent approaches and measures introduced in \cite{curiel2016measure}. 
In this paper, we consider the general model considered in a previous work \cite{d2019copula} and we present a software that permits the calculation of the inequality in a general system composed by a number of interacting individuals. Any individual moves among several communities in time and according to its membership, and depending on that of the other individuals, produces an attribute. The dynamic of individuals among the communities is described according to a piecewise homogeneous Markov chain which requires the identification of an unknown number of changing-points (i.e. where the Markov chain changes its dynamic). Conditional on the occupancy of the communities, the individuals produce an attribute in quantities expressed by a multivariate probability distribution where the dependence structure is managed by a copula function. Finally, using a Monte Carlo algorithm, we show how to compute the moments of the Random Entropy.      

The main innovation brought by this research is the building of the software {\bf{Randentropy}}. It contemplates different aspects that were only partially considered in other research papers. Indeed, different studies deal with software and packages related to multi-state models of Markovian type. For example, in \cite{ferguson2012mssurv} the authors consider a package for computing marginal and conditional occupation probabilities for Markov and non-Markov multi-state models, including the censoring problem and the use of covariates. In \cite{jackson2011multi}, multi-state models for panel data observed continuously and generally based on the Markov assumption have been instead considered. The possibility to obtain a time-varying model is considered using piecewise-constant time-dependent covariates.
Contrarily to these studies, our software gives different transition probability matrices according to the change-points detection methodology presented in \cite{polansky2007detecting}, which is based only on observations of the Markov process and not on additional covariates. 
Moreover, once the piecewise homogeneous Markov chain is identified, the software provides sequences of dependent random vectors denoting the ownership of an attribute by the individuals of the system. Thus, the system becomes a multivariate Markov reward process on which the Random Entropy is evaluated. To our knowledge, our software is the sole that computes the Random Entropy and does it in a very general framework that encompasses 
recent contributions presenting diversity measurement based on (deterministic) entropy where the migration of individuals among the communities is not allowed, see \cite{marcon2015entropart}. Of potential interest is also the use of the software {\bf{Randentropy}} to problems approached with the traditional concept of entropy, see e.g. \cite{behrendt2019rtransferentropy} and \cite{saad2019pymaxent}.\\
The subsequent sections of this paper present the general mathematical model, relevant scenarios of application and the software main characteristics, both the CLI (Command Line Interface) and GUI (Graphical User Interface) are described.

\section{Theory}

The main function driving the development of the software we are presenting here (i.e. {\bf{Randentropy}}) refers to the computation of  a  measure of inequality on the distribution of a given attribute among a set of $N$ individuals. 
 The quantity of this attribute depends on a discriminatory criterion, according to whom the individual  belongs to a given group. 
Accordingly to the nomenclature mainly derived within the ecology community, but preserving its general validity also in other domains, we denote the set of  individuals as a meta-community that is partitioned in several interacting groups called communities. This description is the same adopted in \cite{entropart}.

Let denote the meta-community by $\mathcal{C}$ and the number of its members by $N$. Each individual $c$ $ \in \mathcal{C}$  belongs, at any time $t \in \N$, to one of $D$ different communities that form the meta-community $\mathcal{C}$. The variable $x^{c}(t)$ with values in ${E} =\{1 , 2 , ...,  D \}$ denotes the community to which the individual $c$ belongs to at time $t$. Every time the individual is a member of a given community, it owns a quantity of the personal attribute denoted by $s^c(t)$.
The considered system is stochastic, in the sense that each individual passes through different communities randomly in the course of time and, as a consequence, the personal attributes evolve over time randomly. In this way, the proposed approach is more general as compared to that proposed by \cite{entropart}, where the possibility for members to migrate from a community to another is not permitted. 

The sequence of the visited communities by any individual $c \in \mathcal{C}$, that is $\{x^{c}(t)\}_{t\in \N}$, is assumed to be a realization of a stochastic processes $\textbf{X}^c := (X^c(t))_{t \in \N}$. Thus, the sequences of individual's attribute, that is $\{s^{c}(t)\}_{t\in \N}$, evolve randomly too. We will denote, from now on, the stochastic process describing the evolution of individuals' attribute as $\cs^c:=(S^c(t))_{t \in \N}$. 
The processes $\textbf{X}^c$ and $\textbf{S}^c$ evolve jointly, meaning that: the evolution of the process $\textbf{S}^c$ is driven by the stochastic process $\textbf{X}^c$, which controls it. 
A precise description of this mechanism follows.\\
Firstly, we assume an independence assumption between the dynamics of the individuals. Thus, the community process for every individual will be denoted simply by $\X = \X(t)$, and the reference to specific individual $ c \in \mathcal{C}$ is dropped.\\ 
Moreover, we assume that $\X = \X(t)$ is distributed according to a piecewise homogeneous Markov chain (PHMC).  
The process $\X$ is a PHMC taking values in the  finite set $E$, if  it exists a positive number of change-points
$k$, a sequence $\tau_0 = 0 < \dots < \tau_k = \infty$ of increasing times and a sequence ${}^{(0)} \bfP, \dots, {}^{(k)}\bfP$ of stochastic matrices (such that for any $l \in \N, l \leq k$), it ensues that: for any $t \in \{\tau_l,\dots, \tau_{l+1}-1\}$ and any $i,j  \in$ E the following Markov property holds:

\begin{equation*}
\begin{aligned}
    & \P \left( X(t+1) = j | X(t) = i,  X(0:(t-1)) = i_{0:(t-1)} \right) \\
    & = \P \left( X(t+1) = j | X(t) = i \right) = \,^{(l)}p_{ij}.
\end{aligned}
\end{equation*}

The symbols  $i_{0:{(t-1)}} = (i_0, \dots, i_{t-1}) \in E^t$, $X(0:(t-1)) = (X(0), \dots, X(t-1))$  and $ \{\tau_l,\dots, \tau_{l+1}-1\}$  represents the time interval, enclosed between  the $l^{th}$  and the $l+1^{th}$ change-point,
 where the dynamics at community-level are fixed and described by the transition probability matrix  $ \,^{(l)}\bfP=\{\, ^{(l)}p_{ij}\}_{i,j \in E}$.\\
Intuitively, the term piecewise refers to the existence of some points in time where the dynamic change consistently. These times are called change-points. They break up the time-line into several sub-periods within whom the Markov process is homogeneous.\\

Next step concerns the specification of the processes describing the personal attributes, i.e. $\textbf{S}^c$.
We consider a meta-community where the personal attributes of the individuals can be considered to be dependent among each others.  The dependence is introduced through the application of a copula function. 
This strategy is pursued assuming that the marginal distributions of the attributes of the individuals allocated in the same community, and they share a common probability distribution. Formally, let $ F_x$ denote the conditional distribution of attribute  $S^c(t)$ knowing  the community $X^c(t) =x$ of the individual $c\in \mathcal{C}$, i.e. 

$$
 F_x := \mathcal{D}(S^c(t) | X^c(t) =x), \quad \textrm{for any} \quad t \in \N, 
$$

where, for a given random variable $A$, the symbol $\mathcal{D}(A)$ denotes its probability distribution.\\
Now we are in the position to advance the second main assumption stating that: the conditional joint distribution of $(S^1(t), \dots, S^N(t))$ knowing $(X^1(t) = x^1, \dots, X^N(t) = x^N)$ is given by

\begin{equation*}
    \begin{aligned}
&\mathcal{D}(S^1(t), \dots, S^N(t) | X^1(t) = x^1, \dots, X^N(t) = x^N)\\
& =C_\theta (F_{x^1}, \dots, F_{x^N}),
\end{aligned}
\end{equation*}

where $C_\theta$ is the  copula, with dependence parameter $\theta$. According to the considered copula function, $\theta$ may also be a vector of parameters.\\ 

As we are interested in measuring the inequality of the distribution of attributes in the meta-community, we need to introduce a measure of inequality. In particular, the measure of inequality we consider allows the user to face with stochastic processes. 
The measure is based on 
the Theil entropy (see \cite{theil}), closely related to the Shannon entropy (see \cite{shannon}). 
Given a probability distribution 
\[
\textbf{p} = (p^1, \dots, p^N),\,\,\,\,p^{i}\geq 0,\,\,\, \sum_{i=1}^{N}p^{i}=1,
\]
the Theil index, $T(\textbf{p})$ of $\textbf{p}$, is defined as the Kullback-Leibler (KL) divergence $\mathbb{K}(\textbf{p}|\textbf{u})$ between $\textbf{p}$ and the uniform distribution $\textbf{u}$, or equivalently, as the difference between $\log(N)$ and the Shannon entropy $S({\textbf{p}})$. Precisely,
\begin{equation}\label{EqnStaticTheilIndex}
T(\textbf{p}) := \mathbb{K}(\textbf{p}|\textbf{u}) := \sum_{i=1}^N p^i \cdot \log(N \cdot p^i) = \log(N) - S(\textbf{p}), 
\end{equation}
where $S(\textbf{p}) = -\sum_{i = 1}^N p^i \log p^i$.

The definition of Theil index has been extended for stochastic processes by \cite{d2010generalized} and successively applied and further investigated in \cite{d2012} and in \cite{d2014} for an additive decomposition of this index. 
The random extension of the Theil index is, indeed, introduced. 

Let $sh^c(t)$ be the share of the attribute held by individual $c \in \mathcal{C}$ at time $t \in \N$. It is defined as the proportion of its own attribute $S^c(t)$ relative to the sum of the attribute over all individuals 
i.e.,  
\begin{displaymath}
 sh^c(t) =  \frac{S^c(t)}{ \displaystyle \sum_{d \in \mathcal{C}} S^d(t)}. 
\end{displaymath}
The vector of shares of attributes  at time $t$, $sh(t) := (sh^c(t))_{c \in \mathcal{C}}$ defines a probability distribution on the set of countries $\mathcal{C}$. 
 Note that $\textbf{sh} := (sh(t)_{t \in \N})$ is a stochastic process that depends on  the stochastic processes $\textbf{S}^c$, controlled by $\X^c$.
We call Random Entropy of personal attributes in the meta-community the stochastic process $\D$, given by
\begin{equation} \label{EqnDynamicTheilIndex}
\D = \sum_{c \in \mathcal{C}} sh^c(t) \log(N \cdot sh^c(t)), \quad t \in \N.
\end{equation}
An explicit formula for the expected value of $\D$ has been provided in \cite{d2019copula}. Nevertheless, that formula can only be effectively implemented for small sized meta-communities and number of communities. In the contrary case, a Monte Carlo simulation approach can be successfully implemented. The proposed algorithm simulates repeatedly the trajectories of all individuals according to the underlying Markov model, providing the sequence of communities to which each individual belongs in time. Moreover, the personal attributes are simulated by using the copula function with marginal distribution for each individual dependent on the community of membership. The expected value of the Random Entropy can be estimated by averaging, for each time, over all simulated attributes in the meta-community.\\

The algorithm (see Algorithm \ref{mcalgo}) is made of several steps, clearly e are omitting some preliminary tasks, such as: the identification of the number $K$ and dislocation in time of the changing points $\{\tau_{k}\}_{k=1}^{K}$; the corresponding estimation of the transition probability matrices $ \,^{(l)}\bfP=\{\, ^{(l)}p_{ij}\}_{i,j \in E}$; the cdf's $\{F_{x}, x\in E\}$ of attribute depending on the community $x$; and the identificability of the copula function $C_\theta$. Obviously, the software {\bf{Randentropy}} is designed to solve all the aforementioned tasks, included the implementation of the Monte Carlo algorithm which represents the very last step of the computation.\\

For easiness of notation we adopt the following vectorial notation along the Algorithm \ref{mcalgo}:

\begin{itemize}
    \item  $X(t,c)=X^c(t)$ denotes the community to which the individual $c$ belongs to at time $t$. Thus, $X(\cdot,\cdot)$ is a matrix whose values are element of $E$. Its i-th row $X(i,\cdot)$ provides the meta-community configuration at time $i$, that is the allocation of the individuals at that time among the communities. Instead, the j-th column of the matrix ($X(\cdot,j)$) gives the trajectory of the individual $j$ in time, that is, the sequence of communities it visited in time; 
    \item $s(t,c)=sh^{c}(t)$ denotes the share of attribute held by individual $c$ a time $h$. Thus, $s(\cdot,\cdot)$ is a matrix whose values are non-negative real numbers. Its i-th row $s(i,\cdot)$ provides the share of the attribute own by the individuals of the meta-community at time $i$; it represents a probability distribution. The j-th column of the matrix $s(\cdot,j)$ shows instead the evolution in time of the share of the attribute own by the individual $j$;
    \item $DT(t)=\D$ denotes the value of the Random Entropy at time $t$ in the meta-community. In exact term, it gives the Theil's entropy computed on the probability distribution $s(t,\cdot)$ which represents a realization of the Random Entropy in the given simulation;
    \item $ M$ denotes the horizon time of the simulation.\\

\end{itemize}

\newcommand{\INDSTATE}[1][1]{\STATE\hspace{#1\algorithmicindent}}

\begin{algorithm}[H]
\begin{algorithmic}
\INDSTATE[2]   for $c=1:N$ \\
\INDSTATE[4]      set $X(0,c)=i_{c}$; \\
\INDSTATE[2]  end for \\
\INDSTATE[2]  set $k=1$; \\
\INDSTATE[2] 3.\,  set $h=\tau_{k-1}$ while $h<(\tau_{k}\wedge M)$ \\
\INDSTATE[4]     for c=1:N \\ 
\INDSTATE[6]         sample the random variable $X\sim \, ^{(k)}p_{X(h-1,c),\cdot}$ \\
\INDSTATE[6]        set $X(h,c)=X(\omega)$; \\
\INDSTATE[4]     end for \\
\INDSTATE[4]     sample $(v_{1},v_{2},\ldots,v_{N})$ from $N$ independent Uniform $U(0,1)$;\\
\INDSTATE[4]     set $u_{1}=v_{1}$ and $S(h,1)=F_{X(h,1)}^{-1}(u_{1})$;\\
\INDSTATE[4]     for $b=2:N$\\
\INDSTATE[6]       set $C_{\theta}(v_{b}|(u_{1},\ldots,u_{b-1}))=\frac{\partial_{(u_{1},\ldots,u_{b-1})}^{b-1}C_{\theta}(u_{1},\ldots,u_{b-1},v_{b},1,\ldots,1)}{\partial_{(u_{1},\ldots,u_{b-1})}^{b-1}C_{\theta}(u_{1},\ldots,u_{b-1},1,1,\ldots,1)}$;\\
\INDSTATE[6]       set $u_{b}=C_{\theta}^{-1}(v_{b}|(u_{1},\ldots,u_{b-1}))$;\\
\INDSTATE[6]       set $S(h,b)=F_{X(h,b)}^{-1}(u_{b})$;\\
\INDSTATE[6]       set $s(h,b)=\frac{S(h,b)}{\sum_{b=1}^{N}S(h.b)}$;\\
\INDSTATE[4]     end for\\
\INDSTATE[4]    set $s(h,1)=1-\sum_{c=2}^{N}s(h,c)$;\\
\INDSTATE[4]     set $DT(h)=\sum_{c=1}^{N}s(h,c)\cdot \log(N \cdot s(h,c))$;\\
\INDSTATE[4]     set $k=k+1$ and continue to 3.
\INDSTATE[2]  end while \\

\end{algorithmic}
  \caption{Monte Carlo Simulation of the Random Entropy}
\label{mcalgo}
\end{algorithm}

The result of Algorithm \ref{mcalgo} is a sequence of values $\{DT(h)\}$, $\,h=1,\ldots,M$. Now, if we execute the cited algorithm $L$ times, we can denote by $\{DT^{(l)}(h)\},\,h=1,\ldots,M$ the result of the simulation at the l-th repetition. Then, we are able to provide an estimation of the expected value of the Random Entropy by the average value in the $L$ simulation, i.e.
$$
\hat{DT}(h)=\frac{1}{L}\sum_{l=1}^{L}DT^{(l)}(h), \,\,h=1,\ldots,M.
$$

\section{Relevant scenarios of application}

In this section we provide a short description of two possible domains of application of the model. Certainly, a variety of additional situations falls well within the described theoretical setting.\\

\subsection{Financial inequality in an economic area}\label{finappl}

This application was originally considered by \cite{discrete} and \cite{continuous} and successively in a more comprehensive way in \cite{d2019copula}. In this framework we have a meta-community that coincides with a given set of countries all belonging to a given Economic area. A possible case is represented by the European Economic Area. Practically, every country receives a note about its financial creditworthiness, which is expressed in term of a sovereign credit rating, see e.g. \cite{trueck2009rating} and \cite{d2017semi}. Credit ratings are measured in an ordinal scale and assigned by the rating agencies. Moody’s, Standard \& Poor’s and Fitch are the three major among others.\\ Each rating class can be seen as a community, in which the countries are allocated at every time. According to the own riskiness (expressed by rating class), each country pays interest rates on its debt. When the interest rates are compared to a benchmark they define the so-called credit spreads. Thus, credit spreads can be seen as the personal attributes held by each country in time. 

Empirical analysis has shown that credit spreads of European countries are positively correlated, with the exception of Denmark, Sweden and the United Kingdom. To model this complex correlation structure a copula function can be used accordingly to our framework. Once the credit spreads are obtained, it is possible to compute the vector of attributes at time $t$, $sh(t) := (sh^c(t))_{c \in \mathcal{C}}$. Finally, the computation of the expected value of $\D$ gives an effective tool for forecasting the financial inequality in an economic area and its evolution in time. 

\subsection{Human mobility and environmental implications}

Another area in which the $\D$ might be useful is related to the analysis of human mobility data and specific attributes of interest, see e.g. \cite{song2006evaluating} and \cite{krumme2013predictability}. Evidently, it is  possible to use Markov chains as a tool to measure patterns of movements of individuals in a given area. Substantially, the global area, in which the totality of individuals (the meta-community) lives, is partitioned into different locations (communities) and the probability of the next visited location is assumed to depend only on the current location and not on the previous ones. As members of a given location, individuals possess a personal attribute that can be of different nature. 

For example, it would be possible to consider pollution as a variable depending on the specific location, and to measure with the index $\D$ the inequality of the distribution of pollution in the global area and how it may evolve in time. Another possible choice, for the personal attribute, can be the level of expenditures, in such a case the Random Entropy could be used for assessing the inequality of expenditures in the area. 
The latter approach can represent an indeed useful tool to optimize the displacement policies of new markets and stores.

\section{Computational details and applications}

The software we are presenting here has been engineered so that the main computational kernel is included in a single python module 
named {\bf randentropymod} \cite{gitrepo}. The cited module contains two classes: {\bf randentropykernel} and {\bf changepoint}. The two classes  are devoted to the Markov reward approach computation, and to the change-point estimation, respectively.
The full software bundle is then composed by two Command Line Interfaces (CLIs): {\bf randentropy.py} and 
{\bf randentropy\_qt.py}, and a single Graphical User Interface (GUI) based on PyQt5 \cite{gui} (i.e. the Python binding of 
the cross-platform GUI toolkit).

While the two mentioned CLIs, have been specifically developed to perform separately the Markov reward computation 
(i.e. {\bf randentropy.py}) and the change-point estimation (i.e. {\bf changepoint.py}), the GUI has a wider 
ability. Indeed, the GUI may be used to perform both the change-point estimation as well as the Markov reward computation,
and clearly also to easily visualize and explore the results obtained.

The full software suite has been developed within the Linux OS environment. However, once the needed packages 
are downloaded and installed, it should work, without restrictions, also under Mac OS and Windows thanks to the intrinsic portability nature of
the Python programming language. 
The Python packages,  in addition to the aforementioned PyQT5, strictly needed to run the code are: \texttt{Numpy}  (see \cite{numpy}) 
and \texttt{Scipy} (\cite{scipy}) used to engineered the numerical tasks, \texttt{matplotlib} for the plots 
and data visualization (see \cite{matplotlib}). 

\subsection{The  randentropykernelclass and related CLI}
\label{hmc:cli}

As already stated, the {\bf randentropykernel}  class is devoted to the computation of the Random Entropy which is based on the Markov model with dependent rewards as described in Section 2. The class is made 
of several methods as the one to specify the community matrix (i.e. {\bf set\_community}) and the attributes matrix 
(i.e. {\bf set\_attributes}) which correspond to the matrices $X(\cdot, \cdot)$ and $s(\cdot, \cdot)$ used in the algorithm, respectively. There are clearly various methods to tune the computation behavior such as: set the 
number of Monte Carlo simulation steps (i.e. 
{\bf set\_num\_of\_mc\_iterations}), or the simulated time period  
{\bf set\_simulated\_time}. 
Finally, the user has the ability to enable or disable the copula function  
via the {\bf set\_usecopula} method,  and clearly to perform the main computation calling the 
{\bf run\_computation} method. 
Once the computation is completed, the user can retrieve all the results: the first and the second-order moments
of the Random Entropy using {\bf get\_entropy} and {\bf get\_entropy\_sigma}, respectively. 

\begin{figure}[!ht]
\centering
\includegraphics[width=0.5\textwidth]{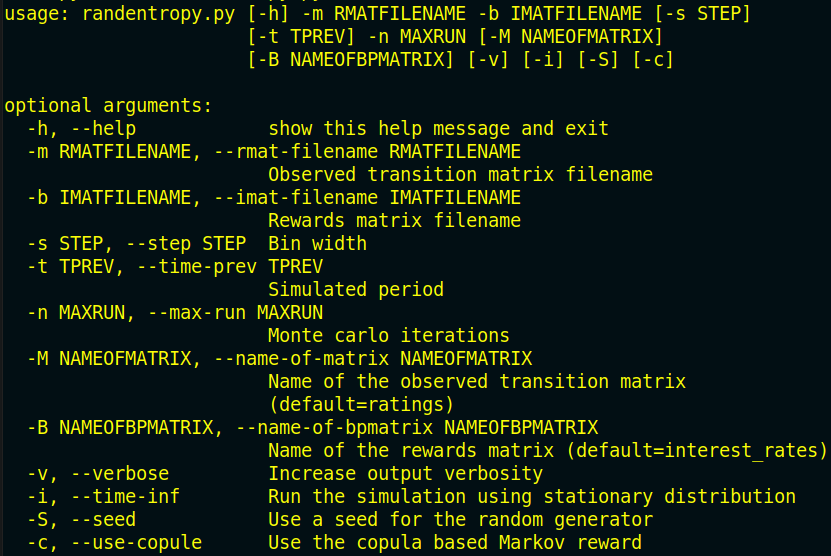}
\caption{CLI for the Markov reward approach}\label{cli:1}
\end{figure}

The {\bf randentropy.py} is the CLI that is naturally bonded to the mentioned class.  As can be seen from Figure (\ref{cli:1}),
the user has the possibility to specify two input matrices (i.e. to specify both their locations and names): the first one representing the community matrix, while the second is the Attributes one. The mentioned matrices may be stored both 
on a MatLab file or on a CSV style one. 

Evidently, the CLI options reported in Figure (\ref{cli:1}) reflect the cited {\bf randentropykernel}  capabilities. Then, \texttt{-s} allows for the 
bin width specification, needed to estimate the probability distribution of the attribute given the community membership. 
Secondly, \texttt{-t} enables the user to specify the simulated period, and \texttt{-n} refers to the number of Monte Carlo iterations. 
Optionally, the \texttt{-i} flag allows the user to run the simulation after computing the stationary distribution. 

It is finally somehow interesting 
to report here that: in case one wants to perform the simulation using the stationary distribution $\pi$ of the Markov chain $\X = \X(t)$ we need to solve a linear matrix equation $a x = b$.  
To solve the given equation one can  compute the value of  $x$  that minimizes the Euclidean 2-norm  $|| b - a x ||^2$. This has been done by applying a specific 
function within Numpy libraries (see \cite{numpy}).

\subsection{ The changepoint class and related CLI}

As already stated within the {\bf randentropymod} module there is also the {\bf changepoint} class. The cited class, and thus the related CLI, is devoted to  
detect the position of $k$ change-points, where $k = 1, 2, 3$. 
In particular, the code finds the positions of the change-points by maximizing the likelihood function of the observed trajectories of the of the members within their communities. At the same time, the $\Lambda$ test 
is carried out in order to assess statistically significant differences among the transition probability matrices found. Additional details on this statistical test are available in \cite{polansky2007detecting} and \cite{d2019copula}.  

The most relevant methods within the class are needed to specify the transition matrix (i.e. {\bf set\_community}) and the number of 
change-points to be detected (i.e. {\bf set\_num\_of\_cps}). Once the initial settings have been specified the main computation starts using 
the {\bf compute\_cps} method. Finally, the calculated $x$ change-points can be 
retrieved using the {\bf get\_cp1\_found} , {\bf get\_cp2\_found} and {\bf get\_cp3\_found}  respectively for the first, second and third change-point.

\begin{figure}[!ht]
\centering
\includegraphics[width=0.5\textwidth]{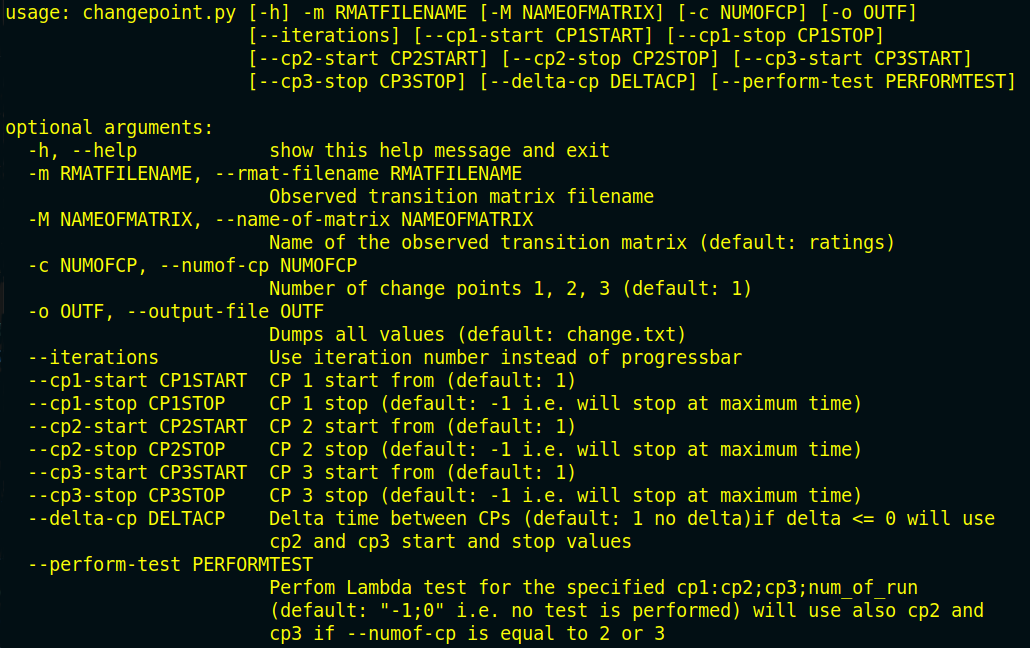}
\caption{CLI for change-point detection algorithm}\label{cli:2}
\end{figure}

Once again the CLI options, reported in Figure \ref{cli:2}, as expected, reflect the class capabilities.
Thus, to run the code, the input transition matrix has to be specified, in terms of a Matlab or a CSV filename, as well as the matrix name within the file (options \texttt{-m} and \texttt{-M}, respectively). The number of change-points to be considered has to be defined as well (i.e. using the \texttt{-c} option), otherwise the code will run assuming a single change-point.
Optionally, an output filename, where all the results are written, can be specified using the \texttt{-o\/--output-file} option.

Finally, we introduced some methods, and clearly the 
relatives CLI options, that can be used also to distribute the
computational burden among several processes, thus CPUs.
Indeed, while working with a huge amount of data 
it can be convenient to specify a range of time within whom the
algorithm is carried out, or to use a specific time distance  
between two change-points.
Thus, the user has the ability to define a range of time for the first change-point (the same apply for the others)
via the \texttt{set\_cp1\_start\_stop} method. Similarly, using the 
\texttt{set\_delta\_cp} method, one can specify the delta time to be considered  among the change-points.

\subsection{Graphical User Interface}

All the previously illustrated functionalities, have been integrated also on a GUI (Graphical User Interface). The GUI has been implemented using PyQT5 a comprehensive set of Python bindings for Qt v5 \cite{pyqt_docu}.
While we implemented two different CLIs, to fully cover the various aspects
implemented within the \texttt{randentropymod}, the GUI is unique and can be access via
the \texttt{randentropy\_qt.py} file \cite{gitrepo}.

\begin{figure}[!ht]
\begin{minipage}{0.49\textwidth}1
\includegraphics[width=0.90\textwidth]{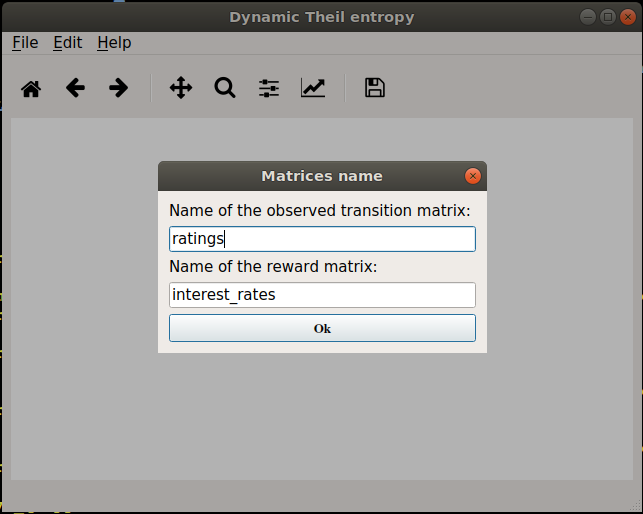}
\caption{Dialog to specify the input matrices}\label{first:interface}
\end{minipage}
\hspace{0.1cm}
\begin{minipage}{0.49\textwidth}
\includegraphics[width=0.90\textwidth]{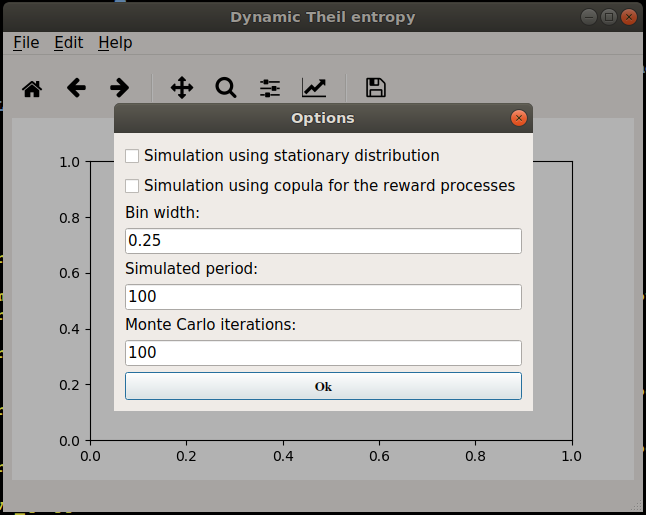}
\caption{Dialog to specify the the input parameters related to the Monte-Carlo simulation}\label{mc:option}
\end{minipage}
\end{figure}

The computation starts after choosing an input file, can be both a Matlab as
well as a CSV one,  containing two matrices. 
The first matrix has to contain the data of the variable which is supposed to
evolve according to a Homogeneous Markov Chain (HMC) (e.g. in the financial application the variable consists on the sovereign credit ratings, see Section 
\ref{finappl}). 
As a matter of fact, the first matrix is expected to be named ''ratings'' by default
(see Figure \ref{first:interface}). The second matrix has to refer to the reward process describing the attribute which is driven by the HMC. In the case of the financial application, as illustrated in Section \ref{finappl}, this  
is the credit spread. As the code directly computes the credit spread starting from the interest rates, the second matrix directly collects the 
interest rate data. Indeed, by default, this matrix within the file is expected to be named ''interest\_rates'' (see Figure \ref{first:interface}).

Once the two matrices have been specified, the user may 
start the computation: \texttt{Edit -> Run}. The use is prompted with a dialog 
window, reported in Figure (\ref{mc:option}, where has the ability to specify: 
the bin width to estimate the empirical distributions (one for each ordered variable of the first matrix),
the simulated period and the number of Monte Carlo iterations. 

\begin{figure}[!ht]
\centering
\includegraphics[width=0.50\textwidth]{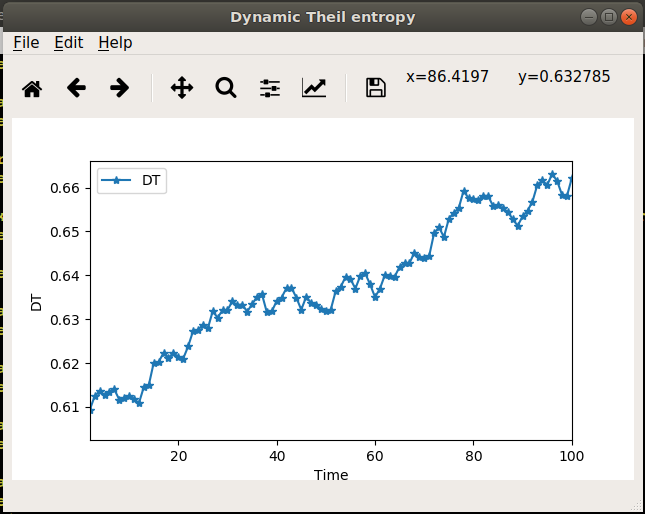}
\caption{Output: dynamic inequality}\label{plot}
\end{figure}

Alternatively, the user can flag ''Simulation using stationary distribution'' to compute the asymptotic values of the Random Theil's Entropy.
After pushing on \texttt{OK} button the program will start the computation, and as finished it returns the plot of the Dynamic inequality (Figure (\ref{plot})),
that the user has the ability to interact with and to save as a
graphical file (i.e. PNG, PDF, PS, and more). 

Subsequently, by clicking on \texttt{Edit -> Plot CS distributions} the user 
can plot the histograms of the empirical distributions of the attribute. 
Moreover by clicking on \texttt{Edit -> View Transition matrix} the transition probability matrix, estimated on the sequences of visited communities, is shown. 

\begin{figure}[!ht]
\begin{minipage}{0.49\textwidth}
\includegraphics[width=0.9\textwidth]{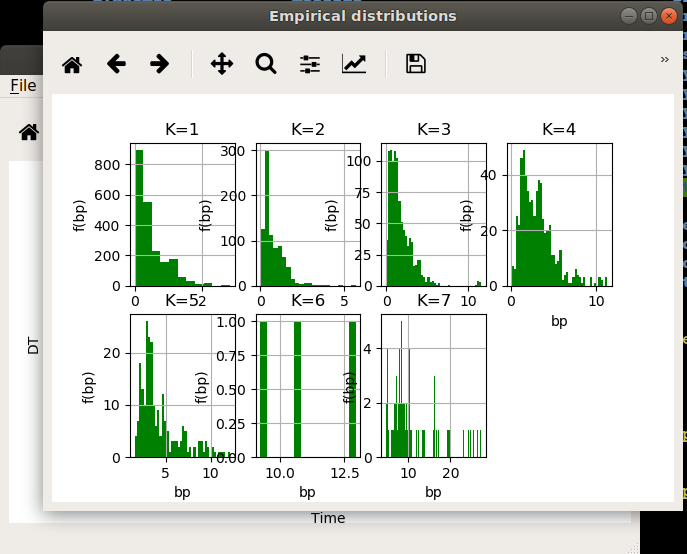}
\end{minipage}
\hspace{0.1cm}
\begin{minipage}{0.49\textwidth}
\includegraphics[width=0.9\textwidth]{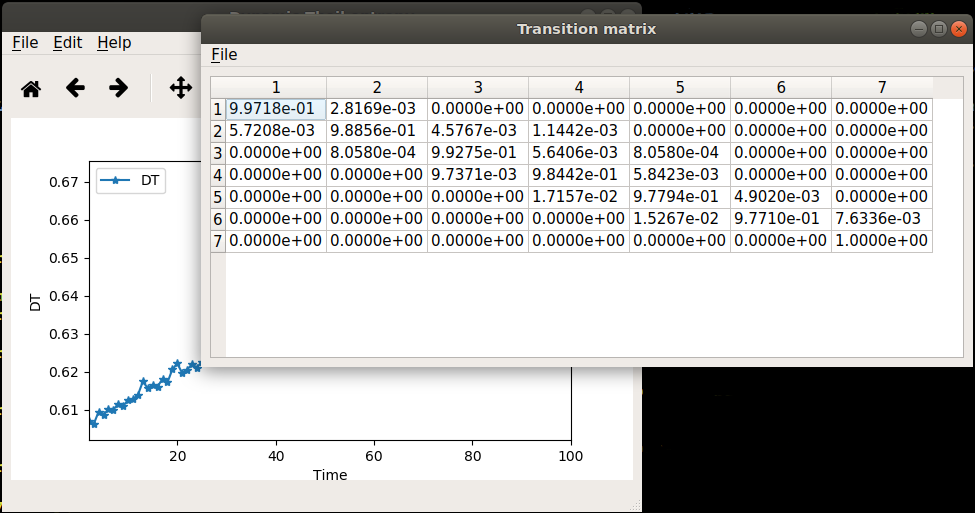}
\end{minipage}
\caption{Histogram of the CS empirical distribution  / Transition probability matrix}
\end{figure}

\begin{figure}[!ht]
\begin{minipage}{0.49\textwidth}
\includegraphics[width=0.9\textwidth]{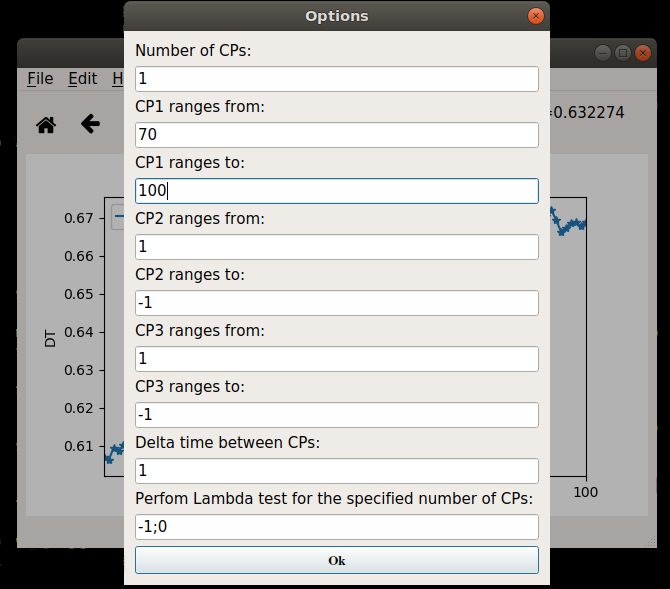}
\end{minipage}
\hspace{0.1cm}
\begin{minipage}{0.49\textwidth}
\includegraphics[width=0.90\textwidth]{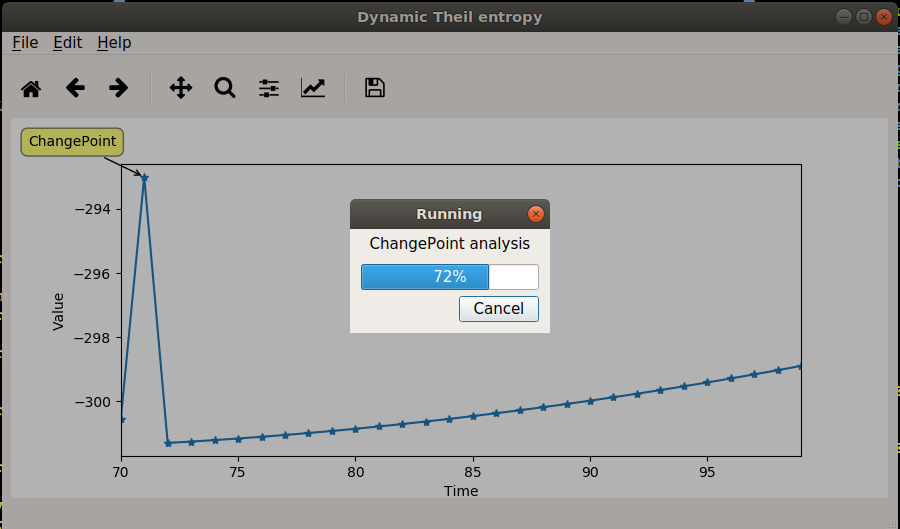}
\end{minipage}
\caption{Options required for change-point detection}\label{option:cp}
\end{figure}

Finally, with \texttt{Edit -> RunChangePoint} one can run the change-point detection algorithm. As  described  for the CLI, the code run after the specification of: the number of change-points to be detected and the corresponding $\Lambda$ test (see Figure \ref{option:cp}); the range of time where the algorithm is carried out and, eventually,  the distance between two subsequent change-points.

In the case reported in Figure (\ref{option:cp}), a single change-point is detected within a range of time spreading between $t=70$ and $t=100$.

\begin{figure}[!ht]
\begin{minipage}{0.49\textwidth}
\includegraphics[width=0.9\textwidth]{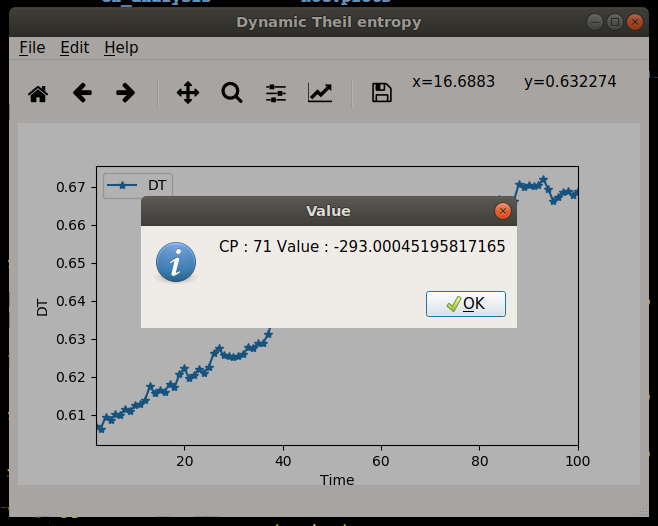}
\end{minipage}
\hspace{0.1cm}
\begin{minipage}{0.49\textwidth}
\includegraphics[width=0.9\textwidth]{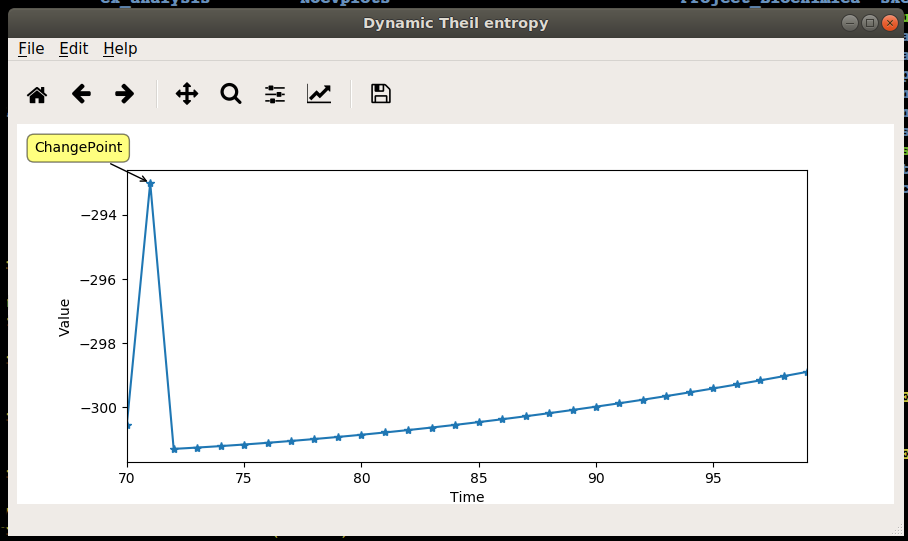}
\end{minipage}
\caption{Output of the change-point detection algorithm}\label{like_cp}
\end{figure}

After confirming the chosen options, the computation starts and the GUI returns the plot of the likelihood function estimated on the community data (see in Figure \ref{like_cp}), together with the value of the maximum likelihood function, and the corresponding position of the calculated change-point. Evidently, also in this case, the resulting plot can be saved into a standard graphical 
file format. 

\subsection{Testing financial inequality in an economic area}

Finally, we will show how the described CLIs and GUI can be used to predict the
financially inequality in the European Economic Area according to the theoretical model proposed in \cite{discrete, continuous}.
In this specific case the meta-community coincides with all the countries within the 
European Community. Thus, each rating class, as assigned by rating agencies,  
can be seen as a community, in which the countries are allocated at every time step.
Clearly, as also already stated in the previous section, the credit spread represents 
the personal attributes held by each country.

The results we are here reporting have been obtained using the monthly rating, attributed by the Standard \& Poor’s agency, to the
26 European countries (UK and Cyprus have been excluded in the current meta-community sample) from January 1998 to 
December 2016 (see \cite{discrete} for extra details on the 
data-set we are here considering). 

To detect the position of a change-point, within the considered horizon time, we compute the maximum value of the likelihood function considered as a function of the position of the changing point. Finally, we fix the change point as the value that maximizes the likelihood function. In the proposed software one can use both the the
{\bf changepoint.py} CLI as well as the  GUI: 

\begin{center}
$ python3 changepoint.py -m ./files/sep_monthly.mat -c 1$
\end{center}

\begin{figure}[!ht]
\centering
\includegraphics[width=0.5\textwidth]{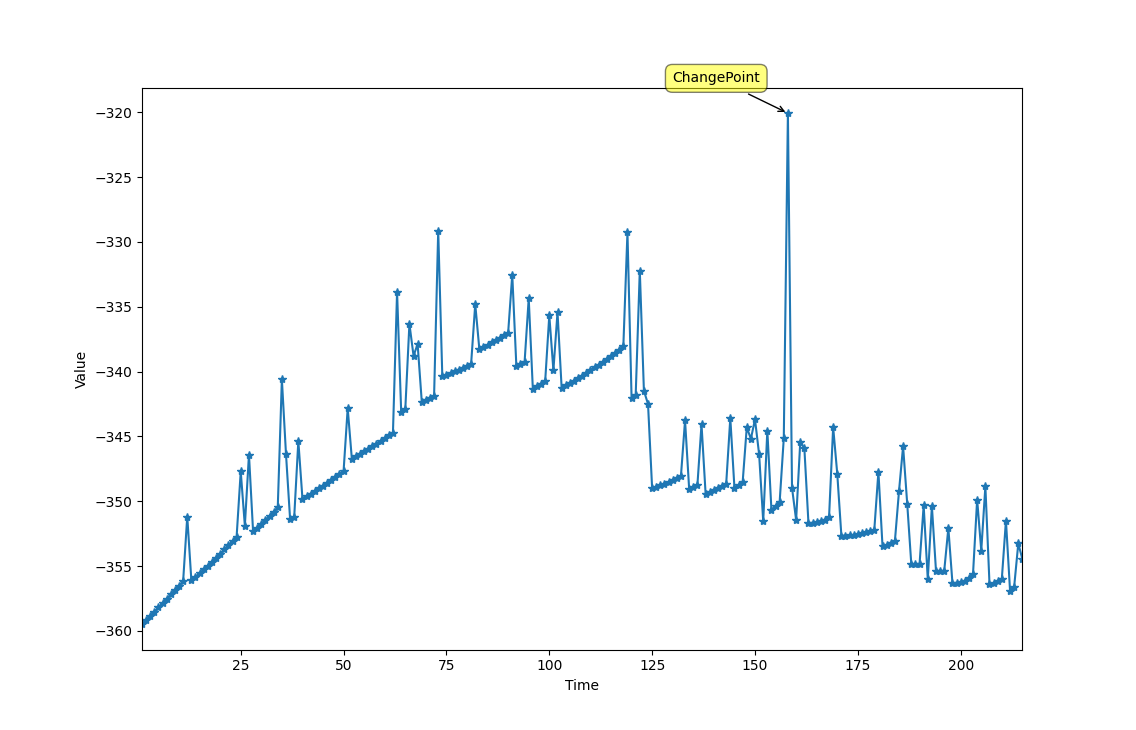}
\caption{GUI results for the change-point detection , see text for details.}\label{changepointdetection}
\end{figure}

The result is reported in Fig. \ref{changepointdetection}, where the likelihood function is computed depending on the position of the change point (measured on the X-axis). The software detects a change-point at time 158 (the maximum value of the likelihood function). The value 158 corresponds to a change point detected in January 2012. Indeed, at the beginning of 2012 the value of the total credit spread in Europe had a peak of about 10.000 basis
points (bp) and this growth was driven by the rise of the securities yield of Greece (2.924 bp), Ireland
(1.245 bp) and Portugal (1.385 bp), see \cite{continuous} for more detail about the evolution of financial variables.

\begin{figure}[!ht]
\begin{minipage}{0.49\textwidth}
\includegraphics[width=0.9\textwidth]{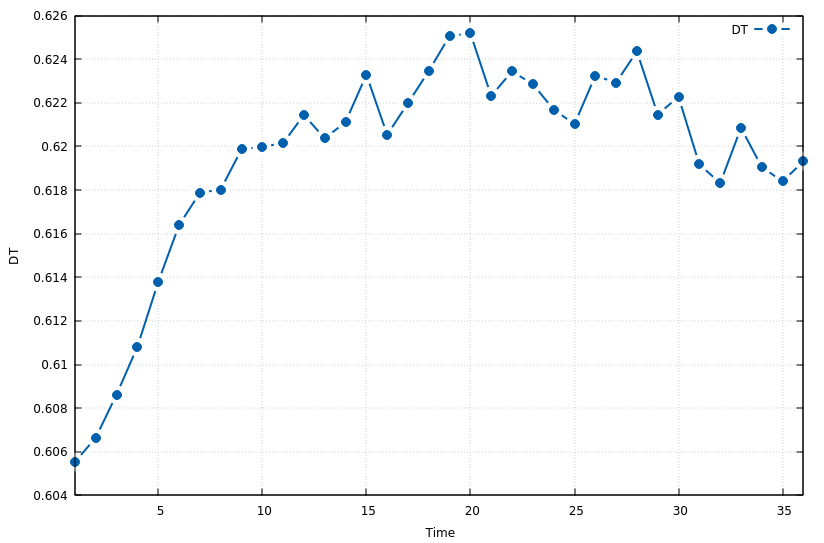}
\end{minipage}
\hspace{0.1cm}
\begin{minipage}{0.49\textwidth}
\includegraphics[width=0.9\textwidth]{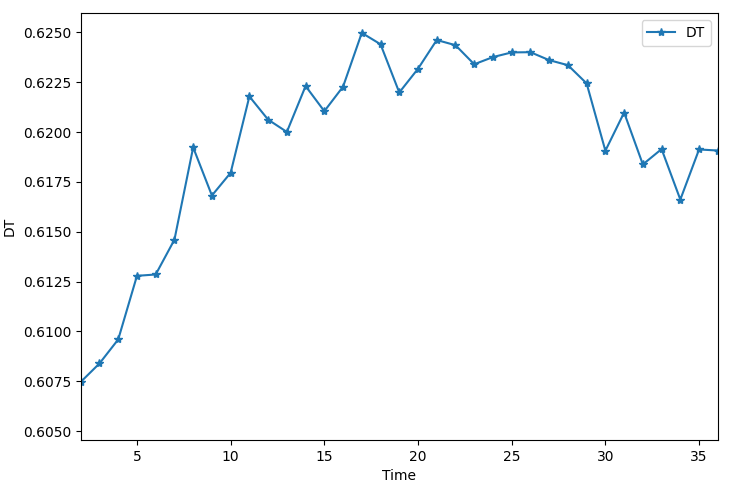}
\end{minipage}
\caption{Random Entropy. Results obtained using the CLI are reported on the left panel, while the ones obtained using the GUI have been reported on the right panel}\label{entroresults}
\end{figure}

Equivalently, a user can forecast the financial inequality in an economic area and its evolution in
time via the {\bf randentropy.py} (or the GUI):

\begin{center}
$ python3 randentropy.py -m ./files/sep_monthly.mat  \ 
     -b ./files/sep_monthly.mat -s 0.25 -t 36 -n 1000 -v $
\end{center}

where the forecast period has been set to 36 months, using 1000 Monte Carlo simulations. 
The final result, reported in Fig. \ref{entroresults}, shows a similar trend for both the GUI and CLI,  
with some clear differences related to the implicit randomness of the Monte Carlo procedure (clearly the user can easily 
avoid this difference selecting a fixed random seed using the --seed option, or equivalently via the  \texttt{set\_use\_a\_seed} method within the \texttt{randentropykernel} class ).
The results entail a sharp increase in short-term financial inequality, as measured in term of credit spread, which is expected to persist in the first 10 months of the forecast. Then, the rise is expected to be less pronounced until the reaching of its maximum value around month 20. Immediately afterwards, a slight decrease is expected to be observed.

As a final remark it is somehow important to underline that, evidently, a user has the capability of 
building its own code, to perform the same or  similar computations just described, accessing directly the 
functionalities implemented within  the {\bf randentropymod} Python 3.x module.

\section{Conclusions and perspectives}

The {\bf{Randentropy}} software allows estimating the inequality in a stochastic system according to the framework based on Random Entropy as developed in \cite{d2019copula}. The methodology is able to consider dependent behaviours of the individuals and time-varying dynamics, which may be of interest in several applied domains.
Possible developments of the research include the possibility to consider semi-Markov models, as done in the {\bf{SemiMarkov}} R Package developed by \cite{krol2015semimarkov}, to which a reward scheme based on a copula function should be attached, followed by the evaluation of the Random Entropy according to our software.\\
\indent Random Entropy evaluation, in the presented general framework, is a new and challenging subject of research and is not available in any software; this renders our investigation an “unicum” in the literature of inequality assessment in stochastic systems.

\bibliography{mybib}
\bibliographystyle{jss}

\end{document}